\documentstyle[epsfig,prl,aps]{revtex}

\begin{document}
\author{I. I. Mazin$^{a,b}$ and David J. Singh$^a$}
\address{$^a$Complex Systems Theory Branch,
Naval Research Laboratory, Washington, DC 20375-5320\\
$^b$Computational Science and Informatics,
George Mason University, Fairfax,
VA}
\title{Ferromagnetic spin fluctuation induced superconductivity in Sr$_2$RuO$_4$}
\draft
\twocolumn[\hsize\textwidth\columnwidth\hsize\csname@twocolumnfalse\endcsname
\maketitle

\begin{abstract}
We propose a quantitative model for triplet superconductivity in Sr$_2$RuO$%
_4 $ based on first principles calculations for the electronic structure and
magnetic susceptibility. The superconductivity is due to ferromagnetic spin
fluctuations, that are strong at small wave vectors. The calculated
effective mass renormalization, renormalized susceptibility, and
superconducting critical temperature are all in good agreement with
experiment. The  order parameters is of
comparable magnitude on all three sheets of the Fermi surface. 
\end{abstract}

]

The layered ruthenate, Sr$_2$RuO$_4$ has attracted considerable recent
interest. It is structurally similar to the first cuprate
superconductor, (La,Sr)$_2$CuO$_4$, is near a magnetic instability 
(Sr$_x$Ca$_{1-x}$RuO$_3$ and Sr$_2$RuYO$_6$ are ferro- and
antiferromagnetic, respectively), and was thought to be strongly correlated.
However, closer examination reveals more and more differences from the
cuprates. It was noted that 
SrRuO$_3$ is ferromagnetic (FM) so it was conjectured that Sr$_2$RuO$_4$ must be
close to a FM instability as well\cite{rice}. This has recently
been corroborated by detailed microscopic calculations of magnetic
properties of ruthenates\cite{MS}. Ferromagnetic fluctuations disfavor both $%
s-$ and $d-$wave superconductivity, so it was suggested\cite{rice,jap} that
superconductivity in Sr$_2$RuO$_4$ must be triplet ($p$-wave), thus very
different from the high T$_c$ cuprates. The idea of the strongly correlated
electrons in Sr$_2$RuO$_4$ is mostly based on the apparent disagreement of
Angular Resolved Photoemission (ARPES) measurements of the Fermi surface\cite
{ARPES} with results of LDA calculations\cite{singh,oguchi}. This argument
is, however, questionable, because ARPES measurements  disagree with de
Haas-van Alphen (dHvA) experiments\cite{dHvA}. The notorious failure of the
LSDA to describe properly the magnetism of undoped cuprates does
not occur in ruthenates\cite{MS}. Thus, the case for the strong correlations
in ruthenates is questionable.

The following challenges should be met by a quantitative theory of the
electronic states and superconductivity in Sr$_2$RuO$_4$: (i) reconciliation
of the (well reproducible) ARPES results with the dHvA measurements. This
may also relate to whether or not there are strong correlation effects. 
(ii) the mechanism for superconductivity and how it is related with
the  large mass renormalization (of a factor of 3-4). In this
Letter we address both of these issues.

\begin{figure}[tbp]
\centerline{\epsfig{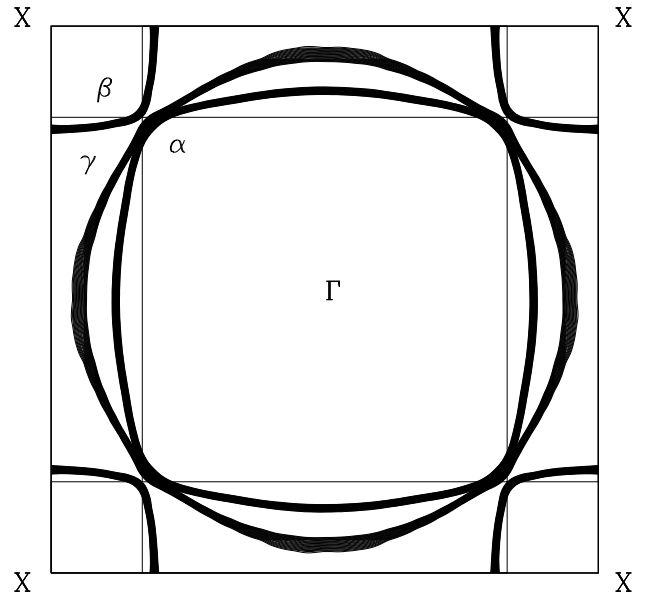}}
\vskip 2mm
\caption{LAPW Fermi surface. The
origin is at $\Gamma$. The thickness of the  lines is inversely
proportional to the Fermi velocity (the inner and the outer contours
are $E_F\pm 2$ mRy). The model Fermi surfaces $\xi$ and $\zeta$ are
shown as straight lines. The model Fermi surface $\gamma$ is within the 
2 mRy window around the actual surface and thus not shown.}\label{FS}
\end{figure}

The valence bands of Sr$_{2}$RuO$_{4}$ are formed by the three $t_{2g}$ Ru
orbitals, $xy,$ $yz,$ and $zx.$ These are hybridized with the in-plane
oxygen and, to a lesser extent, with the apical oxygen\cite{singh,oguchi} $p$%
-states. The bare oxygen $p$ levels are well ($\sim 2$ eV) removed from 
$E_F$, so the effect of the O $p$ orbital is chiefly renormalization
of the ionic $t_{2g}$ levels, and assisting in the $d-d$ hopping. The LDA
band structure can be reasonably well described in the vicinity of the Fermi
level as three mutually non-hybridizing tight-binding bands: $\epsilon _{xy}(%
{\bf k)}=E_{0}+2t_{dd\pi }(\cos ak_{x}+\cos ak_{y})+4t_{dd\pi }^{\prime
}\cos ak_{x}\cos ak_{y},$ and $\epsilon _{{%
{x \atopwithdelims\{\} y}%
}z}({\bf k)}=E_{0}+2t_{dd\pi }{%
{\cos ak_{x} \atopwithdelims\{\} \cos ak_{y}}%
}+8t_{\perp }\cos{ak_{x}\over 2}
\cos{ak_{y}\over 2}\cos{ak_{z}\over 2},$ with the parameter ($%
E_{0}-E_{F},t_{dd\pi },t_{dd\pi }^{\prime },t_{\perp })$ being (-0.4, 0.4,
-0.12,0) and (-0.3, 0.25, 0, -0.025) eV for the $xy$ and $xz,$ $yz$ bands,
respectively, for the bands of Ref.\cite{singh}. With nearest neighbors
only, this model yields one nearly circular cylindrical electronic sheet ($%
\gamma )$ of the Fermi surface (FS) and four crossing planes (quasi-1D 
FS). The weak $xz-yz$ hybridization reconnects these planes to form
two tetragonal prisms, a hole one ($\alpha )$ and an electron one ($\beta )$%
, as  in Fig.\ref{FS}. ARPES gives a FS of different topology: the van Hove
singularity at {\bf k}$=(\pi /a,0),$ which appears slightly above the Fermi
level in the calculations ($\approx 60$ meV in the LDA calculations and $%
\approx 50$ meV with the gradient correction\cite{GGA} included), is seen
below it in photoemission experiments. This reconnects the surface $\gamma $
and makes it hole-like instead of electron-like. The total electron count in
the ARPES FS is still 4, indicating stoichiometry of the samples. Important
consequences were ascribed to the fact that the van Hove singularity is
situated in the same place as in the cuprates.
The main difference between the LDA and the ARPES Fermi surfaces is that the
latter corresponds to a larger $E_0({%
{x \atopwithdelims\{\} y}%
}z)-E(xy)$ (the $d_{{%
{x \atopwithdelims\{\} y}%
}z}$ levels are higher because of an additional hybridization with the
apical (O2) oxygen $p_{{%
{x \atopwithdelims\{\} y}%
}}$ orbitals). Importantly, the calculations imply strong Stoner
renormalization. The Stoner factor $I,$ calculated as described in Ref. \cite
{MS}, is 0.43 eV, and $N(0)=2.06$ eV$^{-1}$\cite{singh}. This
yields a Stoner renormalization $1/(1-IN)=9,$ somewhat larger than deduced
from experimental susceptibility\cite{chi}, $\chi /\chi _{band}=7.3.$ 
To fit the
experiment, $I$ should be $I_{\exp }=0.42$ eV. Note that the experiment
leaves no room for any renormalization of $\chi $ beyond the Stoner one.

However, the topology of the ARPES FS disagrees with that from dHvA
experiments. The latter yields three cross-sections, which sum up to 4
electrons/cell with excellent accuracy {\it only if the surface }$\gamma $ 
{\it is electron-like.} The LDA calculated $\alpha ,$ $\beta ,$ and $\gamma $
areas deviate from the dHvA experiment by only -2\%, -3\% and 5\% of the
Brillouin zone area, respectively, and an exact match can be achieved by
very slight shifts of the bands $\alpha ,$ $\beta ,$ and $\gamma $ by 5, -4,
and -3 mRy, respectively. Such agreement is generally considered very good
even in simple metals, and the small mismatch (which does not change the FS
topology) is likely due to some underestimation in  LDA calculations of
the tiny $xz-yz$ hybridization. Unlike dHvA, which probes the bulk, ARPES
probes essentially first surface Ru-O layer. The cleavage plane in Sr$_{2}$%
RuO$_{4}$ is likely associated with the rocksalt layers, leaving the
Ru-O2 bond dangling or otherwise strongly perturbed. As such, this bond is
likely to be contracted compared to the bulk, and the electronic structure
of the surface RuO$_{2}$ layer differs from 
 bulk. The main effect of such a surface relaxation is expected to be a
strong modification of the Ru(${%
{x \atopwithdelims\{\} y}%
}z)$-O2(${%
{x \atopwithdelims\{\} y}%
})$ hopping. In a linear approximation, this can be estimated from
bulk calculations with the Ru-O2 bond length reduced by a half of the
supposed surface contraction of this bond. We performed such calculations
for Sr$_{2}$RuO$_{4}$ with the O2 shifted 
by 0.1 \AA\  and found that the
energy distance between the Fermi level and the van Hove singularity was
reduced by 30 meV. Thus, the surface relaxation of the Ru-O2 bond that would
bring our LAPW calculations in agreement with the ARPES measurements would
be less than 0.4 \AA\ (probably, closer to 0.3 {\AA }, due to non-linearity%
\cite{nonl}). Although the actual surface relaxation for Sr$_{2}$RuO$_{4}$
is not known, the change in the observed electronic structure due to the
Ru-O2 bond relaxation is in the right direction and of the right order of
magnitude compared to the observed ARPES FS. We conclude that the LDA and
dHvA yield the bulk electronic structure of Sr$_{2}$RuO$_{4}.$ The
differences in ARPES  presumably reflect the surface structure.

This said, we  recall that the mass renormalization (1+$\lambda )$ found
in dHvA experiments\cite{dHvA} and from the specific heat\cite{sheat} is
unusually large: for the $xy$ ($\gamma )$ sheet it is 4, and 3.3 for the two
other sheets. Materials with an electron-phonon coupling constant of the
order of 2.5 are known, but if it were so large in Sr$_{2}$RuO$_{4},$ with
its high phonon frequencies, the superconducting $T_{c}$ would be much
higher than 1.5 K. This paradox is
naturally resolved in the framework of the conjecture\cite{rice,jap} that Sr$%
_{2}$RuO$_{4}$ has strong electron-paramagnon coupling, and may even
be a $p$-wave superconductor, which is also in accord with recent
experiments showing anomalously strong dependence of $T_{c}$ on residual
resistivity\cite{maenoAPS}. In such a case, two different coupling constants
appear: $\lambda _{0}^{m}$ which controls the mass renormalization is the
average of the electron-paramagnon interaction over the FS, while $\lambda
_{1}^{m}$ which determines the $p$-wave transition temperature is the $l=1$
angular component of this interaction. Importantly, this holds for any
boson-mediated interaction, including electron-phonon, so that $\lambda
_{0}=\lambda _{0}^{m}+\lambda _{0}^{p}$, $\lambda _{1}=\lambda
_{1}^{m}+\lambda _{1}^{p}.$ For $s$-pairing although the mass
renormalization is controlled by $\lambda _{0}^{m}+\lambda _{0}^{p},$ the
superconducting coupling constant is $\lambda _{0}^{p}-\lambda _{0}^{m}.$

The situation with Sr$_2$RuO$_4$ is further complicated by the fact that
there are three different sheets of the FS and the order parameters on all
sheets should be determined simultaneously. Without a quantitative numerical
estimate it is impossible to assess whether or not the triplet pairing
hypothesis of Refs.\cite{rice,jap} can be reconciled with the body of
experimental facts. Fortunately, the LSDA calculations 
provide the necessary information for a quantitative
analysis.

The most important (and  most uncertain) part of such an analysis is the
interaction responsible for pairing and for the mass renormalization. This
was not specified in the previous works, but we conjecture that it is the
exchange of paramagnons. Such an interaction in metals was studied 
with respect to possible superconductivity in Pd in the late 70-ties (see,
e.g.,\cite{appel,AM}), and later in connection with heavy fermions. 
The parallel-spin interaction, relevant for
triplet pairing is given in the RPA by the sum of the diagrams with  odd
numbers of loops, 
\begin{equation}
V({\bf q=k-k}^{\prime })=\frac{I^{2}(q)\chi _{0}(q)}{1-I^{2}(q)\chi
_{0}^{2}(q)}.  \label{V}
\end{equation}
The mass renormalization is not as easy to define. Besides the 
parallel-spin interaction (\ref{V}), there is the antiparallel-spin
interaction, given in the same approximation by the sum of the
chain diagrams with even numbers of loops, plus ladder diagrams
\cite{appel,BE}. In case of contact interaction, the
total interaction is three times stronger than the interaction in
the parallel-spin channel only.
It was pointed out\cite{AM},
though, that there is no  good physical
reason to single out any particular class of diagrams.
It was found that including all three classes  above leads
to systematic overestimation of mass renormalization by a factor of
2 to 3 \cite{appel,Levin}. 
Our case is further complicated 
because unlike the
electron-phonon interaction, the electron-electron (and, correspondingly,
the electron-paramagnon) interaction is already included in some average way
in the LSDA band structure. Thus, the electron-paramagnon mass renormalization
is to some extent included in the LDA mass as well.

Despite all this difficulties, one can get an idea about the size of
the electron-paramagnon mass renormalization by making calculations
with the parallel-spin interaction (\ref{V}) only; one may consider
that as a lower bound for the total spin-fluctuation induced 
renormalization.
The mass renormalization then 
is computed in the same way as the
electron-phonon renormalization, {\it i.e}, by taking the average of $V({\bf %
q)}$  of Eq.(\ref{V}) over the FS. One has to remember, though, that
there are other effects beyond LDA, apart from the one that we calculate,
which may further increase the observable mass.

The triplet
pairing constant is calculated by averaging $V({\bf q)}$ with the functions
reflecting the {\bf k}-dependence of the direction of the (vector) order
parameter, in the simplest case with ${\bf k\cdot k}^{\prime }/kk^{\prime }.$
A common approximation, which we use here (although it may be not as good in
Sr$_{2}$RuO$_{4}$ as in Pd) is to take $\chi _{0}(q)=\chi _{0}(0)=N.$ The $q$
dependence of $I$ cannot, however, be neglected and has to be specified.
Essentially, it tells us how much the FM state is favored over
AFM states\cite{note}. As discussed in Ref.\cite{MS}, what
favors ferromagnetism over antiferromagnetism in ruthenates is the oxygen
contribution to the Stoner factor. This is determined 
from the band structure calculations as follows: Atomic Stoner factors
for Ru and O ions are calculated in a standard way and are  $I_{Ru}
\approx	0.7$ eV, $I_O\approx 1.6$ eV. The total Stoner factor for the compound
is $I=I_{Ru}\nu _{Ru}^2+2I_O\nu _O^2$, where $\nu _{Ru}$ and
$\nu _O$ are partial densities of states at $E_F$ of Ru and in-plane oxygen;
the contribution of the apical oxygen is negligible. For AFM
ordering, the second term in the expression for $I$ falls out, because
oxygen is nonmagnetic by symmetry. We found the AFM Stoner
factor $I$ for 
Sr$_{2}$RuO$_{4}$ to be smaller than FM one by 14\% (oxygen
contribution $\Delta I=0.06$ eV). A $q$%
-dependence that reflects this effect is $I(q)=I/(1+b^{2}q^{2}),$ where $%
b^{2}=0.5(a/\pi )^{2}\Delta I/(I-\Delta I)\approx 0.08(a/\pi )^{2}
.$ In the following
we use this  $I(q)$ together with Eq.\ref{V}.

Let us now make link to the real FS. In Refs.\cite{rice,rice2} the maximum
full cylindrical symmetry was assumed for all three FS's. This approximation
completely neglects the quasi-1D character of the $xz$ and $yz$ bands and
cannot be used for quantitative purposes. Instead, we retain the cylindrical
approximation for the $xy$ FS $\gamma $ and use the 1D approximation for the 
$xz$ and $yz$ FS's. Then we have three 2D Fermi lines: $\gamma ,$ a circle
with the radius $g\approx 0.9\pi /a,$ $\xi ,$ two lines parallel to $x$ at $%
\approx \pm 2g/3$ from the ${\bf \Gamma }$ point, and $\zeta ,$ the two
corresponding lines parallel to $y$ (Fig.\ref{FS}). Using the standard
multiband
technique\cite{allenB} we now introduce the coupling matrix $\Lambda
_{ij}^s=N^{-1}\sum_{{\bf kk}^{\prime }}\delta (\epsilon _{{\bf k}i})\delta
(\epsilon _{{\bf k}^{\prime }j})V({\bf k-k}^{\prime })=N\nu _i\nu _j<V({\bf %
k-k}^{\prime })>_{ij},$ where $(i,j)$ can be $\gamma ,$ $\xi ,$ or $\zeta ,$
and $\nu _i=N_i/N$ (from our band structure $v_{F\gamma
}\approx v_{F\xi ,,\zeta },$ and $\nu _\gamma =0.44,$ $\nu _{\xi ,,\zeta
}=0.28).$ Then the mass renormalization in  band $i$ is defined as $%
\lambda _i^s=\nu _i^{-1}\sum_j\Lambda _{ij}.$ The average mass
renormalization is $\lambda ^s=$ $%
\sum_{ij}\Lambda _{ij}^s.$

Using this model, we arrive at  $\Lambda _{\gamma
\gamma }^{s}=0.35,$ $\Lambda _{\xi \xi }^{s}=0.32,$ $\Lambda _{\gamma \xi
}^{s}=0.16,$ $\Lambda _{\xi \zeta }^{s}=0.03.$ This gives $\lambda _{\gamma
}^{s}=(\Lambda _{\gamma \gamma }^{s}+2\Lambda _{\gamma \xi }^{s})/\nu
_{\gamma }=1.5,$ $\lambda _{\xi }^{s}=(\Lambda _{\xi \xi }^{s}+\Lambda
_{\gamma \xi }^{s}+\Lambda _{\xi \zeta }^{s})/\nu _{\xi }=1.8,$ $\lambda
^{s}=1.7,$ to be compared with experimental dHvA values of 3, 2.3, and 3,
respectively. The difference may be due to an electron-phonon coupling of
the order of 1 and/or antiparallel spin fluctuations, neglected
in our calculations.

Let us now return to the question of the $p$-wave superconductivity. The
theory for a cylindrical FS is presented exceedingly well by Sigrist et al%
\cite{rice,sigrist,rice2} and need not  be repeated here. The only
difference for a FS of arbitrary shape is that
instead of the ${\bf k}$-vector components, we have to use Allen's FS
 harmonics\cite{allenB}. So, there are four possible unitary planar
states,  all degenerate if spin-orbit is neglected. 
Let us consider, for instance, the $A_{1u}$ state: 
\begin{equation}
{\bf d}_{{\bf k}}=d\frac{{\bf v}_{{\bf k}}}{v_{{\bf k}}},  \label{d}
\end{equation}
where  {\bf v}$_{{\bf k}}$ is the Fermi velocity. This state has, generally
speaking, a finite superconducting gap, and thus zero density of states at
the Fermi energy below $T_{c},$ in contrast with the experiment\cite{sheat}.
The same holds for three other states, degenerate with the one of Eq.(\ref{d}%
). Nonunitary linear combinations of the states which are gapless are
also possible. These have, however, generally speaking, lower pairing energy
and should not occur.

We now calculate the transition temperature within our spin-fluctuation
model. Similar to Agterberg {\it et al}\cite{rice2}, we consider the
superconducting state with the order parameter $d$ which is constant for
each of the three FS sheets, but differing between the sheets. We have 
to calculate the  matrix $\Lambda _{ij}^{p}=N\nu _{i}\nu _{j}<V({\bf %
k-k}^{\prime })({\bf d}_{{\bf k}}^{i}\cdot {\bf d}_{{\bf k}^{\prime
}}^{j})/(d_{{\bf k}}^{i}d_{{\bf k}^{\prime }}^{j})>_{ij},$ where $i$ and $j$
label the three bands, and find the maximum eigenvalue of the matrix $\nu
_{i}^{-1}\Lambda _{ij}^{p}$\cite{allenB}. The corresponding eigenvector 
defines the coefficient $a$ and thus the relative magnitude of the order
parameter in bands $\gamma $ and ($\xi ,\zeta )$. By symmetry, the pairing
matrix looks like 
\[
\left( 
\begin{array}{ccc}
\Lambda _{\gamma \gamma }^{p} & \Lambda _{\gamma \xi }^{p} & \Lambda
_{\gamma \xi }^{p} \\ 
\Lambda _{\gamma \xi }^{p} & \Lambda _{\xi \xi }^{p} & 0 \\ 
\Lambda _{\gamma \xi }^{p} & 0 & \Lambda _{\xi \xi }^{p}
\end{array}
\right) 
\]
(If we had used instead of $\xi ,\zeta $  nomenclature the $\alpha ,\beta $
one, as in Ref.\cite{rice2}, this symmetry would not hold). Numerical
calculations give $\Lambda _{\gamma \gamma }^{p}=0.16,$ $\Lambda _{\xi \xi
}^{p}=0.075,$ and $\Lambda _{\gamma \xi }^{p}=0.025.$ The maximum eigenvalue
of the corresponding coupling matrix is 0.43, and the corresponding
superconducting state is $0.85\gamma +0.38\xi +0.38\zeta .$

Let us now estimate the transition temperature. Using the characteristic
paramagnon energy $\omega _{sf}\sim (N^{-1}-I)/4\approx 160$ K from our
calculations, as the cut-off frequency, and the Allen-Dynes formula for
strong coupling (although $\lambda =0.43$ is relatively weak, the relevant
number is the renormalization parameter $\lambda ^s\approx 1.7),$ we obtain 
\begin{equation}
T_c\approx \frac{(N^{-1}-I)/4}{1.2k_B}\exp [-(1+\lambda ^s)/\lambda
_{eff}^p]=0.25{\rm K.}  \label{Tc}
\end{equation}
Again, as in the case of mass renormalization, there is some room for the
electron-phonon coupling as well.

One of the key problems, as discussed in Refs. \cite{sigrist,rice2}, is the
residual electronic specific heat\cite{sheat}, which remains at about 50\%
of its normal value in the superconducting regime. This led Agterberg {\it %
et al} \cite{rice2} to postulate a pairing matrix which yields a vanishing
gap for the $\gamma $ band. This, however, does not square with
the quantitative estimate presented in this Letter. An earlier assumption%
\cite{sigrist,jap} was that the excess pairing energy that forbids
nonunitary combination of the order parameters (\ref{d}) may be overcome by
additional magnetic (Stoner) energy in a nonunitary state. The requirements are
 strong Stoner renormalization ( supported by
the calculations) and strong particle hole asymmetry\cite{ferro}. However, a
quantitative estimate according to Ref.\cite{ferro}
shows that the effect is by far too weak. The criterion is $\left[ \frac{%
T_{c}d\log N}{dE_{F}}\right] ^{2}\frac{1}{1-IN}\log \frac{\omega _{sf}}{T_{c}%
}\sim 10^{-5},$ while it should be of the order 1 for the nonunitary state to
stabilize. So, the problem of the residual electronic specific heat remains
open, although it cannot be excluded that it results from sample
inhomogeneity and is extrinsic to superconductivity.

To summarize, we have presented first principles calculations which indicate
that (1) conventional LDA calculations give a correct description of the
bulk electronic structure of Sr$_{2}$RuO$_{4},$ as well as of its
renormalization due to the Stoner exchange interaction; (2) the difference
between the bulk and the surface electronic structure (as measured by ARPES)
can be explained by the surface effect; (3)
interactions due to exchange of FM spin fluctuations, as
calculated from the LDA band structure, are sufficiently strong to explain
both the mass renormalization and superconducting critical temperature.
We would like to emphasize the main approximations
used: (a) Neglect of the {\bf q}-dependence of $\chi
_0$, (b) neglect of strong coupling effects beyond the  Allen-Dynes formula 
(note that strong coupling effect will lead to a final density of states
$N(0)$ at all $T\neq 0$ and thus to nonexponential $T$-dependencies of specific
heat and like quantities), and (c) neglect of correlation effects in 
the
mass renormalization beyond the parallel-spin paramagnon induced interaction.
In principle, it is clear how to improve the first two items, while
the last issue lacks full theory and cannot be easily dealt with.

We acknowledge helpful discussions with W. Pickett, R. Rudd, and D.
Hess. Work at NRL is supported by the ONR. Computations
were performed at the DoD HPCMO NAVO Center.

\begin{references}
\bibitem{nonl}  Since oxygen $p$ bands are well below $E_F$, the
effect is proportional to $t^{2}$, where $t$ is the $pd\pi $ Ru-O2 hopping%
, for which the canonical scaling is $t\propto d^{-4}$.
\bibitem{note}  We do not discuss here the interplanar magnetic
interactions, which, contrary to the intraplanar one, are weak and probably
antiferromagnetic\cite{MS}.
\end{references}
\end{document}